\documentclass[12pt]{article}
\usepackage[utf8]{inputenc}
\usepackage[export]{adjustbox}
\usepackage{authblk}
\usepackage{pgfpages}
\usepackage{graphicx}
\usepackage{multicol}
\usepackage{csquotes}
\usepackage{multirow}
\usepackage{graphicx} 
\usepackage{booktabs} 
\usepackage{authblk}
\usepackage{caption}
\usepackage{subcaption}
\usepackage{stackengine}
\usepackage{amsmath}
\usepackage[english]{babel}

\usepackage{amssymb}
\usepackage{stackrel}
\newcommand{\NYT}{\text{NYT}}
\newcommand{\FFR}{\text{FFR}}
\newcommand{\FOMC}{\text{FOMC}}

\usepackage[a4paper, margin=1in]{geometry}

\usepackage{hyperref}
\hypersetup{
    colorlinks=true,
    linkcolor=blue,
    filecolor=blue,      
    urlcolor=blue,
    citecolor=blue,
}

\usepackage{natbib} 

\title{It's not \textit{always} about money, \\
\textit{sometimes} it's about sending a message \\ \medskip \medskip \medskip \large Evidence of Informational Content in Monetary Policy Announcements \normalsize}

\author[1]{Yong Cai}

\author[2]{Santiago Camara}

\author[3]{Nicholas Capel}

\affil[1]{Northwestern University}
\affil[2]{Northwestern University \& Red-NIE}
\affil[3]{Amazon}

\date{\today}

\begin{document}

\maketitle

\begin{abstract}
    
    This paper introduces a transparent framework to identify the informational content of FOMC announcements. We do so by modelling the expectations of the FOMC and private sector agents using state of the art computational linguistic tools on both FOMC statements and New York Times articles. We identify the informational content of FOMC announcements as the projection of high frequency movements in financial assets onto differences in expectations. Our recovered series is intuitively reasonable and shows that information disclosure has a significant impact on the yields of short-term government bonds.
    
    \medskip
    
\end{abstract}


\section{Introduction} \label{sec:introduction}

The identification and macroeconomic impact of monetary policy shocks is a longstanding and open question. The key challenge is that monetary policy responds to both the current and the expected future states of the macroeconomy. Controlling for these unobserved state variables underlies the main difficulty in identifying the causal impact of monetary policy. One way to overcome this challenge, is to measure changes in the prices of financial assets around Central Bank announcements. Under tight time-windows around Central Bank's announcements, it is credible to assume that financial markets react to said announcements only. However, as stressed by \cite{nakamura2018high} and \cite{jarocinski2020deconstructing} among others, these announcements reveal both policy decisions and the central bank's assessment about the present and future economic outlook. Consequently, these high-frequency movements in the prices of financial assets are a response not only to policy decisions but also to the private sector's updated beliefs about the state of the economy. Consider the following example of both policy and private market beliefs around a FOMC meeting. 

On September 18, 2007, the FOMC cut the Federal Funds Rate by 50 basis points. As the New York Times reports, ``while an interest rate cut was widely expected, there had been profound uncertainty about whether the Fed would choose a more cautious quarter-point reduction. But the bolder action and an accompanying statement, both approved by a unanimous vote of the central bank’s policy-setting committee, made it clear that the Fed had decided the risks of a recession were too big to ignore." Hence, we see how investors learnt from the aggressive policy decision about the seriously negative prospects facing the economy. 

In order to study the causal effects of monetary policy, we would like to decompose high-frequency movements into the part which responds to monetary policy (the monetary component) and the part which responds to information disclosure by the central bank (the information or news component). Studying the effect of the monetary component would allow us to understand the effect of monetary policy, while the information component allows us to understand how central banks influence the economy beyond direct policy tools. 

In this paper, we use FOMC statements and newspaper articles to identify and quantify the information component of monetary policy announcements. We do so via state-of-the art text analysis tools in a framework that is both transparent and highly interpretable. Whereas earlier literature has inferred the information component through the introduction of a structural model (see \cite{nakamura2018high}) or imposing restrictions on the moments of high-frequency surprises (see \cite{jarocinski2020deconstructing}), our method does not impose any assumptions on the relationship between the potential shocks and macroeconomic variables. Instead, we use text data to model expectations of the FOMC and the private sector over the Federal Funds Rate (FFR). Our key insight is that the informational component of monetary policy is the part of high frequency movements that is correlated with differences in the expectations of the two.

Our method consists of three stages. First, we use tools in computational linguistics to embed FOMC statements and New York Times articles as numerical vectors. Next, we use the embedded documents to separately model both FOMC and private sector expectations of the FFR via the elastic net, a transparent method for dealing with high dimensional data, and which \cite{giannone2021sparsity} advocates for economic data. Here, we use the expectation of the FOMC to refer to the hypothetical monetary policy rate that would have been chosen in the absence of monetary policy shocks. Stage two allows us to construct the differences in expectation about the FFR which arose due to differences in information alone. Finally, we project high frequency movements of select interest rates around FOMC announcements onto the differences in expectation and interpret the projection as the information component of the high frequency movements. 

We apply our framework to a set of identified high-frequency movements which the literature has interpreted as monetary policy shocks.
Following \cite{nakamura2018high} and \cite{jarocinski2020deconstructing} we decompose these high-frequency movements into a monetary component and news component. We find that over our sample period of January 2000 to March 2014, differences in beliefs have predictive power over the high-frequency movement of the Federal Funds Futures around FOMC meetings. We interpret this finding as evidence that FOMC announcements disclose information about the state of the economy.

We show that across these different high-frequency movements, the isolated news component has predictive power over the daily change of short term nominal and real rates around FOMC meetings. In particular, we argue that not controlling for the news component of high-frequency surprises may lead to biases on the impact of monetary policy shocks on domestic financial conditions.  

Our contribution to the literature is three-fold. First, we use newspaper text to proxy for private sector information about monetary policy around FOMC meetings. While text analysis has been used to study the information effects of FOMC policy announcements, existing papers have focused on FOMC statements or minutes exclusively (\cite{lucca2009measuring}, \cite{hansen2018transparency}, \cite{benchimol2020communication}, \cite{handlan2020text}). We argue that any informational effects has to operate through differences in expectations, so that both FOMC and private beliefs must be taken into consideration. Hence we augment our analysis of FOMC statements with New York Times articles. 

Second, we propose a text analysis framework that is interpretable and highly transparent. Our framework uses the black-box output of text analysis methods to construct conditional expectations of various macroeconomic variables. Studying the effects of these variables -- or their differences -- on high frequency interest rate movements allows us to understand the transmission mechanisms of informational effects. Our method is also highly transparent: the elastic net, which is a form of penalized linear regression, performs explicit regularization using only two tuning parameters. 

Third, the empirical applications of our framework shed light on the impact of the informational content of FOMC announcements on both financial and macroeconomic variables. Specifically, the news component of high-frequency monetary policy shocks have a significant impact on short run nominal and real rates, while having little to no effect on longer maturities. Further studying these shocks can provide insights on the specific information advantage the Federal Reserve has over private sector agents. 


The paper is organized as follows. Section \ref{sec:data} describes the data. Section \ref{sec:methodology} lays out the text analysis framework in detail. Section \ref{sec:main_results} provides both qualitative and quantitative evidence that our procedure recovers informational content. Section \ref{sec:conclusion} concludes. 





\section{Data} \label{sec:data}

This section of the paper describes the two main types of data used: (i) text coming from both the Federal Reserve's FOMC statements and New York Times' articles; (ii) high-frequency financial surprises around FOMC announcements.

The sample period is January 2000 to March 2014. The FOMC statement announcements were sourced from the Federal Reserve's website.\footnote{\url{https://www.federalreserve.gov/monetarypolicy/fomc_historical_year.htm}} The FOMC statements are short summaries of the topics discussed by the committee, released following a meeting. This include the assessment of the current economic situation, and policy decisions regarding the target federal funds rate, discount rate, and since 2008 different types of unconventional monetary policy. In January 2000, the Committee announced that it would issue a statement following each regularly scheduled meeting, regardless of whether there had been a change in monetary policy. We drop unscheduled FOMC meetings from our sample as in \cite{nakamura2018high} and \cite{handlan2020text}. In general, an unscheduled meeting reflects highly unusual economic or political circumstances.


The New York Times' articles were sourced from Northwestern Library, which granted us access up to March 2014.\footnote{We accessed and downloaded these articles on April 2 2021.} Our sample consists of articles published on the day of and day prior to the FOMC meetings. Furthermore we exploit the fact that articles are tagged by subjects and restrict our sample to only those which contain the subject ``federal reserve". These articles typically discuss the current economic situations and contain several private sector projections on both the economy and FOMC policy decisions. Thus, they provide us with a proxy for private sector expectations on the FOMC announcement day.

\newpage

We were able to find at least one such article during the day of and the day before each scheduled FOMC meeting date. This is consistent with the idea that the FOMC meetings are an important event for the financial markets such that they receive news coverage in the lead-up. 

The empirical section of the paper uses both high-frequency movements in financial assets and macroeconomic variables for the US. The two high-frequency movements studied are the Fed Funds Futures surprises constructed by \cite{gertler2015monetary} and the policy news shocks constructed by \cite{nakamura2018high}. 

We measure the impact of both the isolated monetary and news components using several daily interest rates. The nominal and real Treasury yields are sourced from FRED. Data on macroeconomic aggregates, such as the industrial production index and the consumer price index, are also sourced from FRED. 

\section{Methodology} \label{sec:methodology}

This section of the paper details the three-stage process by which we extract the informational component of the high frequency movements around FOMC announcements. The three-stages can be summarized as follows:
\begin{itemize}
    \item[Stage 1:] Document Embedding. We use state-of-the-art tools in the fields of natural language processing and machine learning to convert text documents in numeric vectors. 
    \item[Stage 2:]  Constructing Proxy Expectations. We use the vector representations from Stage 1 to predict policy decisions of the FOMC. Under a rational expectation assumption, we interpret our estimated functions as expected policy decisions given the information sets of the private sector and the FOMC respectively. 
    \item[Stage 3:]  Extracting Informational Content. Suppose market movements around FOMC announcements is due to informational effects. Then these movements should be proportional to the difference in expectations. We project high frequency movements onto the difference in expectation to recover their informational component.
\end{itemize}
In the following subsections, we provide further details on their implementation, as well as the assumptions that justify our procedure. 

\newpage

Relative to the literature, the main novelty of our approach is in the use of newspaper articles to measure public expectation. Whereas the literature has thus far focused on the FOMC statements alone (see for instance \cite{handlan2020text}, \cite{hansen2016shocking}), we argue below that the information effect should depend on the difference in the expectations of the public and the FOMC, and that neglecting public expectation leads to omitted variable bias. In addition, with the exception of stage 1, which relies on a pre-trained model, the remaining stages involve only simple methods that are highly transparent. Finally, our method of constructing the intermediate expectation term allows us to probe at the mechanism by the information effect operate, as we explore in section \ref{sec:taylorrule}.

\subsection{Stage 1: Document Embedding} \label{subsec:first_stage}

In stage 1, we convert each FOMC statement and NYT article in $768\times1$-vectors using BERT (Bidirectional Encoder Representations from Transformers, \cite{devlin2019bert}). We highlight the challenge of using text data and explain how our BERT overcomes these problems before detailing our procedure. 

Text data is difficult to analyze because of their high-dimensional nature. The simplest approach to deal with text is to just treat them as ``bag-of-words". With such an approach, the analyst would count the number of unique words in each article or document and then use the word counts as covariates for linear regressions or as input into other models. However, such an approach would work poorly in our sample, which has over 8,000 unique words in fewer than 150 observations. Whereas earlier papers have made progress by choosing which words to include (see for instance \cite{lucca2009measuring}), such ad-hoc methods not only have high researcher degree-of-freedom but could also leave information on the table. 

The high dimensionality problem is compounded by the fact that the meanings of words depend on their context. An FOMC statement that is ``worried about inflation not growth" and one is ``worried about growth not inflation" have identical word counts but express contrary ideas.\footnote{A more subtle example of this ``lexical ambiguity" problem is that ``bank" has completely different meanings in ``river bank" and ``investment bank".} The need to take into account sequential information increases the dimension of text data exponentially. Consider a simple exercise in which we count the number of unique 6-word phrases, so that the example phrases above are counted as distinct entries. In a sample with 8,000 unique words, the possible number of unique phrases could be as high as $8,000^6 = 2.6 \times 10^{23}$. It is therefore largely infeasible to work with counts of phrases alone. 

To render text analysis tractable and effective, we turn to BERT -- a state-of-the-art document embedding method that is able to reduce text input into $768\times1$-vectors while preserving sequential information.  Its output captures the meaning of text input very well, as demonstrated by their usefulness for natural language processing tasks. For example, models based on BERT excel in question and answering tasks, in which given a question and a passage from Wikipedia containing the answer, the task is to predict the answer text span in the passage. Succeeding in these tasks all require some ``understanding" of the text input, giving us assurance that the BERT representations capture the meaning of words. 

The key advantage of using BERT is that it is pre-trained on large amounts of text data (BooksCorpus -- 800M words and English Wikipedia -- 2,500M words). Given only 106 observations, it would have been very difficult to train any reasonably flexible text analysis models. For a similar reason, \cite{handlan2020text} relies on the pre-trained XLNet, a minor variation of BERT, for their analysis. 

We note that other papers have employed Latent Dirichlet Allocations (LDA) models on samples of similar sizes [all the Hansen and McMahon papers], even though it is unclear if they are well-behaved given such small samples. In addition, it is known that LDA models have identification issues so that results are driven by the choice of priors even asymptotically (\cite{ke2021text}). Hence, we consider them to be unsuitable for the task at hand. 

We initialized pretrained BERT models from an open source implementation provided by Hugging Face.\footnote{Hugging Face is a large open-source community that hosts pre-trained deep learning models, mainly aimed at NLP. Their state-of-the-art transformer models are used by Google, Facebook, Microsoft and AWS} (https://huggingface.co/bert-base-uncased) To use BERT, we first cleaned the input data by removing numbers, dates as well as common stop-words. Stop-words are a set of words which do not contain information about the content of the articles, such as prepositions and conjunctions. Removing these words ``enriches" the information content in text articles. Here, we relied on the stop-word list compiled by the text analysis package \emph{gensim}. Next, the words were tokenized. A token is the most fundamental object in a chosen language dictionary (it could be a word, character, or subword). We used the default tokenizer provided by Hugging Face for use with the bert-base-uncased model for tokenization. 

Finally, we split each individual document into windows of 256 tokens, with overlaps of 10 tokens between windows. Each window is tokenized separately, and passed through the BERT model to form individual outputs. The embedding vector of the entire document is then taken to be the mean of the BERT class tokens across all of the windowed outputs.

We note that the cleaning process removes all numbers from the text. Consequently, the actual policy decisions as well as numerical projections about the economy are absent from the documents. This gives us assurance that our method is not simply picking up the policy rates or their forecasts, but rather capturing beliefs about the economy.

\subsection{Stage 2: Constructing Proxy Expectations} \label{subsec:second_stage}

In the second stage, we use the document embeddings to construct proxy expectations for the general public and the FOMC. Our starting point is the equations:
\begin{align*}
    \FFR_t & = f^\NYT(X^\NYT_t) + \varepsilon^\NYT_t \\
    \FFR_t & = f^\FOMC(X^\FOMC_t) + \varepsilon^\FOMC_t
\end{align*}
where $X^\NYT_t$ and $X^\FOMC_t$ are the embeddings of NYT articles and FOMC statement at time $t$. We further assume that
\begin{equation} \label{eq:news_exogeneity}
    \mathbb{E}\left[\varepsilon^\NYT_t \, \big| \, X^\NYT_t\right] = 0 \quad , \quad     \mathbb{E}\left[\varepsilon^\FOMC_t \, \big| \, X^\FOMC_t\right] = 0~.
\end{equation}
We interpret the above assumption as follows. Suppose $X^\NYT_t$ captures all information about the economy that is available to the general public at time $t$. Then public expectation about FFR is given by 
\begin{equation*}
    f^\NYT(X^\NYT_t) = \mathbb{E}\left[\FFR_t \, \big| \, X_t^\NYT\right] ~.
\end{equation*}
The conditional mean assumption
\begin{equation*}
    \mathbb{E}\left[\varepsilon^\NYT_t \, \big| \, X^\NYT_t\right] = 0
\end{equation*}
is thus a rational expectation assumption -- the general public has beliefs that are not systematically biased and are correct on average. 

We assume the same is true for the FOMC. That is, we assume that $X^\FOMC_t$ contains all the information that the FOMC has about the economy at time $t$. $f^\FOMC(X^\FOMC_t)$ is then the Taylor rule that maps FOMC information into interest rate decisions. $\varepsilon^\FOMC_t$ is monetary policy shock -- implementation error that prevents the FOMC from achieving its desired optimal policy rate. We assume that this monetary policy shock is $0$ on average and mean independent from $X_t^\FOMC$. This is consistent with the current views on monetary policy shock, as discussed in \cite{ramey2016macroeconomic}. 

Given our limited sample size, we impose further structure for estimation by assuming that $f^\NYT$ and $f^\FOMC$ take the following linear form:
\begin{align*}
    f^\NYT(X_t^\NYT) & = \alpha^\NYT + \beta^\NYT{}'X^\NYT_t \\
    f^\NYT(X_t^\NYT) & = \alpha^\FOMC + \beta^\FOMC{}'X^\FOMC_t
\end{align*}
Because we use the pre-trained BERT, $X_t^\NYT, X_t^\FOMC \in \mathbf{R}^{768}$. In order to estimate $\alpha$ and $\beta$ from our sample of size 106, we use the elastic net, defined as:
\begin{equation*}
    \left(\hat{\beta}^\NYT, \hat{\alpha}^\NYT \right) = \underset{\beta, \alpha}{\arg\min} \left\lVert \FFR - \alpha - \beta'X^\NYT  \right\rVert_2^2 + \gamma\eta \left( \lVert \alpha \rVert_1 + \lVert \beta \rVert_1\right) + \gamma(1-\eta) \left( \lVert \alpha \rVert^2_2 + \lVert \beta \rVert^2_2\right)
\end{equation*}
\begin{equation*}
    \left(\hat{\beta}^\FOMC, \hat{\alpha}^\FOMC \right) = \underset{\beta, \alpha}{\arg\min} \left\lVert \FFR - \alpha - \beta'X^\FOMC  \right\rVert_2^2 + \gamma\eta \left( \lVert \alpha \rVert_1 + \lVert \beta \rVert_1\right) + \gamma(1-\eta) \left( \lVert \alpha \rVert^2_2 + \lVert \beta \rVert^2_2\right)
\end{equation*}
Here, $\FFR$, $X^\NYT$ and $X^\FOMC$ refer to the stacked matrices of $\FFR_t$, $X_t^\NYT$ and $X_t^\FOMC$ respectively.

The elastic net nests two popular methods of penalized regressions. Suppose $\eta = 1$. Then the elastic net reduces to the LASSO estimator. LASSO performs model selection by setting regression coefficients to zero. The resulting output is typically an interpretable model in which only a few coefficients are non-zero. On the other hand, setting $\eta = 0$ leads to the ridge estimator, a dense model which assigns highly correlated regressors similar coefficients. Since the elastic net is more general, it is able to achieve better prediction accuracy given properly chosen tuning parameters (\cite{zou2005regularization}). Furthermore, it has a Bayesian interpretation as the posterior mode induced by the spike-and-slab prior of \cite{mitchell1988bayesian} and may be more appropriate for economic data (\cite{giannone2021sparsity}). 

With our conditional mean assumptions, the elastic net is $L_2$ consistent as long as the space of the covariates is sufficiently rich (\cite{de2009elastic}). Since we are interested in prediction and not model selection, we do not require any sparsity assumption.

Each elastic net requires two tuning parameters $(\lambda, \eta)$. Because these parameters are chosen to maximize $R^2$ in stage 3, we discuss their selection in the next subsection. 

\subsection{Stage 3: Extracting Informational Content} \label{subsec:third_stage}

In this subsection, we explain how we use the proxy expectations from stage 2 to decompose high frequency movements of interest rates around FOMC announcements into an information and a monetary component.  

Let $\Delta R_t$ denote the high frequency movement of interest rate $R$ around the release of FOMC statements. Suppose $\Delta R_t$ includes reaction by the private sector to information released by the Federal Reserve. This component should be a function to the \emph{difference} in information of the two parties. In particular, if there is no difference in information, this component should be $0$. This motivates the following linear regression:
\begin{equation}\label{eq:stage3}
    \Delta R_t = \zeta + \theta \Delta \mathbb{E}[\FFR_t] + \nu_t \quad , \quad \mathbb{E}\left[\nu_t \, | \, \Delta \mathbb{E}[\FFR_t]\right] = 0,
\end{equation}
where 
\begin{equation*}
    \Delta \mathbb{E}[\FFR_t] := \alpha^\FOMC + \beta^\FOMC{}'X^\FOMC_t  - \alpha^\NYT + \beta^\NYT{}'X^\NYT 
\end{equation*}
is the difference in the expectations of the private sector and the FOMC based solely on difference in information.

For any choice of tuning parameters $(\lambda, \eta)$ in stage 2, we are able to construct
\begin{equation*}
    \widehat{\Delta \mathbb{E}[\FFR_t]} := \hat{\alpha}^\FOMC + \hat{\beta}^\FOMC{}'X^\FOMC_t  - \hat{\alpha}^\NYT + \hat{\beta}^\NYT{}'X^\NYT~.
\end{equation*}
We then choose our tuning parameter to maximise $R^2$ in the regression of $\Delta R_t$ on $\widehat{\Delta \mathbb{E}[\FFR_t]}$, where the latter term is a function of $(\lambda, \eta)$. 

The regression in equation (\ref{eq:stage3}) has clear meaning. Here, we are projecting the high frequency movement onto the differences in proxy expectation and we interpret the projected component as the part of $\Delta R_t$ that is responding to Federal Reserve information.

Suppose we were able exactly capture news component of $\Delta R_t$. We would then be able to interpret the residual in the above regression as true monetary policy shocks. However, we are cautious that much of the news component may remain due in part to the restrictive linear assumption in equation (\ref{eq:stage3}). On the other hand, given the sample size of 106, it is unclear if more flexible functional forms would simply lead to over-fitting. For this reason, the simple linear regression is our preferred specification for stage 3. 

An approach commonly seen in the literature (see for instance \cite{handlan2020text}, \cite{hansen2016shocking}) is to run the regression on the FOMC component only:
\begin{equation}\label{eq:shortregression}
    \Delta R_t = \zeta + \theta \cdot f^\FOMC(X^\FOMC_t) + \xi_t \quad 
\end{equation}
As we argued above, the informational component of the high frequency movement should depend on the difference in information, and not the level directly. Suppose equation \ref{eq:stage3} is true. We can rewrite equation \ref{eq:shortregression} as
\begin{equation*}
    \Delta R_t = \zeta + \theta \cdot f^\FOMC(X^\FOMC_t) - \theta \cdot f^\NYT(X^\NYT_t) + \nu_t \quad , \quad \mathbb{E}\left[\nu_t \, | X^\FOMC_t\right] = 0.
\end{equation*}
Here, 
\begin{equation*}
    \xi_t = - \theta \cdot f^\NYT(X^\NYT_t) + \nu_t
\end{equation*}
is clearly correlated with $f^\FOMC(X^\FOMC_t)$ through $f^\NYT(X^\NYT_t)$, giving rise to omitted variable bias.

Furthermore, equation (\ref{eq:stage3}) is robust in that it can be valid even when the conditions in equation (\ref{eq:news_exogeneity}) is violated.  For example, suppose the FOMC has time varying preferences in trading off inflation and output gap, so that a positive shock is more likely when the FOMC preferences have a hawkish realization. This violates the conditional mean assumption in equation (\ref{eq:shortregression}) as long as the statements contains information about its current hawkishness. However, time-varying preferences, or other omitted variables, do not cause a problem for the regression in (\ref{eq:stage3}) as long as: $$\text{Cov}\left(\nu_t,  f^\FOMC(X^\FOMC_t)\right) = \text{Cov}\left(\nu_t,  f^\NYT(X^\NYT_t)\right)~.$$
Intuitively, our regression is valid as long as these omitted variables skew private sector and FOMC expectations the same way. On the other hand, if equation (\ref{eq:news_exogeneity}) does not hold, \ref{eq:shortregression} will be invalid, even if $X_{t}^\FOMC \perp \!\!\! \perp X_t^\NYT$.

Finally, even if  $X_{t}^\FOMC \perp \!\!\! \perp X_t^\NYT$, our specification will still have more power. Including $f^\NYT(X^\NYT_t)$ would lead to more precise estimates by reducing the variance of the residuals. This gels with the idea that our regression is a more direct test of the information hypothesis: if the high frequency movement is driven by agents updating their information set, theory directly predicts that these movement should be proportional to the amount of updating that occurred. 

Next, we provide intuitive reasons for choosing tuning parameters that maximize $R^2$ in stage 3. Firstly, the tuning parameters that are chosen are unlikely to overfit the stage 2 elastic nets. To see this, observe that if we overfit stage 2 to the extreme, such that $\hat{\alpha}^\FOMC + \hat{\beta}^\FOMC{}'X^\FOMC_t = \hat{\alpha}^\NYT + \hat{\beta}^\NYT{}'X^\NYT_t = \FFR_t$. Then  $\widehat{\Delta \mathbb{E}[\FFR_t]} = 0$ identically so that stage 3 $R^2$ is 0. It is equally unlikely to underfit stage 2. Consider again the extreme case, this time in which $\hat{\beta}^\NYT$ and $\hat{\beta}^\FOMC$ are set to $0$. Now, $\hat{\alpha}^\NYT$ = $\hat{\alpha}^\FOMC = \frac{1}{T} \sum_{t = 1}^T \FFR_t$ so that $\widehat{\Delta \mathbb{E}[\FFR_t]}$ is again $0$. Our procedure guards against both of these problems. Where the method may face issues is in overfitting stage 3. However, we believe that this is less of a concern since stage 3 is a linear regression with only two parameters. 


\subsection{Results}

\begin{table}[htbp]
  \centering
  \caption{Summary Statistics for Stage 3. Robust standard errors are used.}
    \begin{tabular}{ccccccc}
          & Shock & $R^2$ & $\hat{\theta}$ & S.E.  & $t$-stat & $p$-value \\
    \midrule
    \multirow{3}[2]{*}{Pre-ZLB} & PNS   & 0.079 & 0.083 & 0.031 & 2.624 & 0.011 \\
          & FFR   & 0.038 & 0.051 & 0.044 & 1.152 & 0.254 \\
          & FF4   & 0.085 & 0.183 & 0.065 & 2.826 & 0.006 \\
    \midrule
    \multirow{3}[1]{*}{Full Sample} & PNS   & 0.028 & 0.009 & 0.006 & 1.466 & 0.146 \\
          & FFR   & 0.013 & 0.008 & 0.008 & 0.971 & 0.334 \\
          & FF4   & 0.073 & 0.020 & 0.008 & 2.421 & 0.017 \\
    \end{tabular}%
  \label{tab:stage3}%
\end{table}%

Summary statistics for Stage 3 of our procedure is presented in table \ref{tab:stage3}. We find evidence that high frequency movements around FOMC announcements is in part a response to information from the Federal Reserve. In particular, we note $\hat{\beta}$ is positive for PNS, FFR and FF4. It is also significant at the 5\% level for the PNS and FF4 in the pre-ZLB period, and for FF4 in the full sample. The positive sign implies that when the information set of the FOMC leads it to set higher interest rate than the public was expecting, the market response is positive. 

The $R^2$ in stage 3 is relatively low across the board. This could be interpreted as meaning that most of the movement around FOMC announcements is response to pure monetary policy shock. On the other hand, given the our highly restrictive linear model in stage 3, we consider the information effect we found to be a lower bound: with more data and a more flexible stage 3, we might find potentially much larger information effects. 

\begin{table}[htbp]
  \centering
  \caption{Summary Statistics for Stage 2}
    \begin{tabular}{cccccccc}
          & Shock & $\eta^\NYT$ & $\lambda^\NYT$ & NYT $R^2$ & $\eta^\FOMC$ & $\lambda^\FOMC$ & FOMC $R^2$ \\
    \midrule
    \multirow{3}[2]{*}{Pre-ZLB} & PNS   & 0.550 & 1.150 & 0.122 & 0.250 & 3.275 & 0.082 \\
          & FFR   & 0.700 & 0.811 & 0.152 & 0.050 & 13.219 & 0.147 \\
          & FF4   & 0.400 & 1.630 & 0.068 & 0.400 & 2.310 & 0.060 \\
    \midrule
    \multirow{3}[1]{*}{Full Sample} & PNS   & 1.000 & 0.012 & 0.884 & 1.000 & 0.100 & 0.995 \\
          & FFR   & 0.650 & 0.050 & 0.872 & 0.800 & 0.142 & 0.974 \\
          & FF4   & 0.100 & 0.142 & 0.884 & 1.000 & 0.100 & 0.990 \\
    \end{tabular}%
  \label{tab:stage2}%
\end{table}%

Table \ref{tab:stage2} presents summary statistics associated with stage 2, including the tuning parameters chosen by our procedure. Larger $\eta$ implies that the chosen model is closer to a LASSO, while smaller $\eta$ implies that the chosen model is closer to a ridge regression. We see evidence for sparsity in PNS and FF4 with the FOMC statement, especially in the full sample period, since the preferred model is the LASSO. This suggests that there are only a few variables which are important for determining the FOMC's desired policy rate. On the other hand, during the pre-ZLB period, the selected models tended towards ridge regressions, suggesting that there are more diverse considerations for determining interest rates prior to the Great Financial Crisis.

We note that $R^2$ is high in the full sample, but much smaller in the the pre-ZLB sample. This is likely because FFR remained largely constant after the Great Financial Crisis. Taking the model at face value, we would infer that monetary policy is easier to predict after post-crisis, and that the variance of monetary policy shocks is smaller. On the other hand, we are cautious that our linear model may be inappropriate for modelling the policy rates, which may be censored at the zero lower bound. 

We conclude by noting that our approach is highly transparent. All tuning parameters as well as regression parameters that we chose are contained in tables \ref{tab:stage3} and \ref{tab:stage2}. The only unpresented parameters belong to BERT, which is trained by an external party without reference to any of the variables used in stage 2 and 3. 

\subsection{Channels of Information Transmission}\label{sec:taylorrule}

In order to understand how new information from the FOMC translates to market reaction, we consider how differences in expectation of specific macroeconomic variables drive high frequency movements around FOMC announcements. To do that, we extend the analysis thus far to a general economic variable $Y_i$:
\begin{align*}
    Y_{i,t} & = f_i^\NYT(X^\NYT_t) + \varepsilon^\NYT_{i,t} \\
    Y_{i,t} & = f_i^\FOMC(X^\FOMC_t) + \varepsilon^\FOMC_{i,t}
\end{align*}
where as before we assume that:
\begin{equation*}
    \mathbb{E}\left[\varepsilon^\NYT_{i,t} \, \big| \, X^\NYT_t\right] = 0 \quad , \quad     \mathbb{E}\left[\varepsilon^\FOMC_{i,t} \, \big| \, X^\FOMC_t\right] = 0~.
\end{equation*}
However, the stage 3 equation of interest is
\begin{equation}\label{eq:stage3_taylor}
    \Delta R_t = \zeta + \sum_{i=1}^p \theta_i \cdot \Delta \mathbb{E}[Y_{i,t}] + \nu_t \quad , \quad \mathbb{E}\left[\nu_t \, | \, \Delta \mathbb{E}[Y_t]\right] = 0.
\end{equation}
Suppose $Y_{i,t} = \pi_t$ is inflation at $\pi_t$. In the regression above, we would then interpret $\theta_i$ as the high frequency movement that resulted from differences in inflation expectation. Running regression \ref{eq:stage3_taylor} allows us to understand what type of informational differences is driving the high frequency movement.

In particular, let $Y_{1,t}$ be real GDP in the next nearest quarter and $Y_{2,t}$ be CPI in the next nearest month.\footnote{In particular, we use the ``Consumer Price Index: All Items Excluding Food and Energy''. This choice is motivated by the FOMC targetting a core inflation rate which excludes commodity prices such as Food and Energy which are highly influenced by international markets.} These are the two variables that would be relevant if interest rate was set according to a simple Taylor rule. Results are presented in table \ref{tab:stage3_taylorrule}. Across the board, when the FOMC expects higher inflation than the private sector, the interest rates increase following the announcements. This is in line with the predictions of standard macroeconomic theory, in which the central bank raises rates in response to heightened inflationary pressures regardless of whether the source is demand or supply shocks. 

The coefficients on real GDP is mixed. In the pre-ZLB period, the coefficients on real GDP have positive signs across the board. In words, when the FOMC expects higher real GDP than the public, interest rates subsequently increases. In the full sample, however, the sign changes to negative for FFR and FF4. Standard macroeconomic theory does not provide a straightforward answer to how monetary authority should react to GDP fluctuations. For instance, if the central bank receive news of a positive demand shock, then it raises its GDP projections and simultaneously raises interest rates. This gives rise to a positive comovement between expected GDP and monetary rate policy. But, if it receives news of a negative supply shock, a central bank would raise rates even as it lowers GDP projections, leading to negative comovements. Note that the full sample includes episodes of \textit{supply-side} shocks such as tax cuts and/or oil price shocks; and furthermore, we exclude observations from July 2008 to July 2009 inclusive. Hence, it is plausible that in our full sample supply shocks play a large role.

In sum, our method allows us to probe at the mechanisms of information transmission. We find that when the FOMC expects higher inflation than the private sector, interest rates unequivocally increase. However, when the FOMC expects higher real GDP than the private sector, interest rate movements are mixed. These patterns are suggestive of the types of shocks in an economy. 

\begin{table}[htbp]
  \centering
  \caption{Stage 3 Results using real GDP and CPI. Robust standard errors are used.}
    \begin{tabular}{clccccc}
          & \multicolumn{1}{c}{Shock} & $R^2$ & $\hat{\theta}_\text{RGDP}$  & RGDP $p$-value & $\hat{\theta}_\text{CPI}$   & CPI $p$-value \\
    \midrule
    \multirow{3}[2]{*}{Pre-ZLB} & PNS   & 0.103 & 0.001 & 0.073 & 0.005 & 0.131 \\
          & FFR   & 0.083 & 0.000 & 0.378 & -0.080 & 0.256 \\
          & FF4   & 0.216 & 0.030 & 0.000 & 0.032 & 0.440 \\
    \midrule
    \multirow{3}[1]{*}{Full Sample} & PNS   & 0.061 & 0.000 & 0.548 & 0.003 & 0.174 \\
          & FFR   & 0.034 & -0.001 & 0.139 & 0.005 & 0.134 \\
          & FF4   & 0.067 & -0.002 & 0.051 & 0.008 & 0.055 \\
    \end{tabular}%
  \label{tab:stage3_taylorrule}%
\end{table}%

\section{Evidence of Informational Content} \label{sec:main_results}

In this section, we provide evidence that the component we isolated using the procedure in Section \ref{sec:methodology} carries informational content. We focus on event study and financial market evidence.\footnote{In Appendix \ref{subsec:svar_evidence} we present additional evidence by estimating the impact of ``Monetary$_t$'' at the aggregate level using a SVAR model. Expectedly, these are not too different from SVARs using the high frequency shocks alone, given the $R^2$ of Stage 3.}

\subsection{Event Study Analysis} \label{subsec:event_analysis}

We study the components of the high-frequency financial surprises isolated in the previous section. 
\begin{figure}[ht]
    \centering
    \caption{Time Series of PNS$_t$ Decomposition. Following \cite{nakamura2018high} we omit observations between July 2008 to July 2009 inclusive.}
    \label{fig:TimeSeriesPNS}
    \includegraphics[width=12cm,height=9cm]{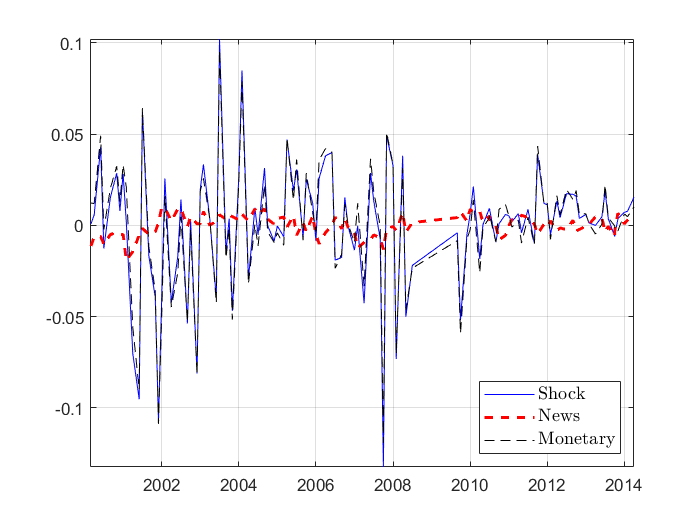}
\end{figure}
Given the modest $R^2$ from the decomposition, it is unsurprising that the news component has relatively smaller variance than the monetary component. However, the series is intuitively reasonable. To show this, we zoom in on three FOMC announcements in which the news component was thought to have played an important role ex post. 

\begin{itemize}
    \item[] \textbf{September 18, 2007}: We return to our opening example on which the FOMC cut its rates by 50 basis points. As our quoted text suggests, the market interpreted the Federal Reserve's fairly aggressive actions and statements to mean that a sharp economic downturn was imminent. Thus it is reassuring that we recorded one of our largest negative news shock on this date. 
    \item[] \textbf{March 18, 2008}: One of our larger positive news shock was documented on March 18, 2008, even amidst the Great Financial Crisis. On this day, the FOMC cut rates by 75 basis points to 2.25 percent. However, as the New York Times reports, ``though it was one of the biggest one-day rate cuts in decades, investors had been betting heavily that the Fed would cut its key rate a full percentage point in response to strong evidence that a recession has begun and to the deepening crisis on Wall Street." Instead, the Federal Reserve policy reflected concerns about inflation. As the FOMC statement writes, ``inflation has been elevated, and some indicators of inflation expectations have risen". It is therefore reassuring that our measure of news recorded a positive value, reflecting the Federal Reserve's information about inflation. 
    \item[] \textbf{January 30, 2002}: Another of our largest positive surprises came as the economy was recovering from the bursting of the dot-com bubble and 9-11. At this meeting, the FOMC left interest rates unchanged, which was a departure from its yearlong policy of rate cuts. According to the FOMC, ``Signs that weakness in demand is abating and economic activity is beginning to firm have become more prevalent". As the New York Times reports, ``Investors interpreted [...] the Fed's decision on interest rates as signs that the economy might be stabilizing. Stocks prices moved higher in the afternoon after the Fed's announcement." Here, we have explicit indication that investors interpreted the Federal Reserve's policy as good news, supporting the positive news shock that we recovered. 
\end{itemize}

\subsection{Financial Market Evidence} \label{subsec:financial_evidence}

The goal of this subsection of the paper is to identify the effect of the informational content of FOMC announcements on nominal and real interest rates of different maturities. 

First, we measure the impact of the informational content of FOMC announcements on nominal and real interest rates. In particular, we estimate the following two empirical equations
\begin{align}
    \Delta r_t &= \alpha_0 + \beta \Delta R_t + e_t \label{eq:shock} \\
    \Delta r_t &= \alpha_1 + \gamma \text{Monetary}_t + \mu \text{News}_t + \epsilon_t \label{eq:shock_decomposition}
\end{align}
were $\Delta r_t$ is the change in the outcome nominal and/or real interest rate of interest, e.g., the Market Yield on U.S. Treasury Securities at 10-Year Constant Maturity; $\Delta R_t$ is the high-frequency financial surprise around the FOMC announcement; $\text{Monetary}_t$ and $\text{News}_t$ are the decomposition coming from the methodology used in Section \ref{sec:methodology}; $e_t$ and $\epsilon_t$ are error terms; and $\alpha_0$, $\alpha_1$, $\beta$, $\gamma$ and $\mu$ are parameters. The parameter of interests are $\beta$, $\gamma$ and $\mu$. Comparing $\beta$ and $\gamma$ allows us to identify how purging the high-frequency surprises of informational content changes the impact of monetary policy shocks on nominal and real interest rates.

First, we study the impact of the high-frequency Fed Funds Futures on nominal and real rates by estimating Equation \ref{eq:shock} using the high-frequency surprise FF4. 
\begin{table}[ht]
    \centering
    \caption{Response of Interest Rates to Fed Funds Futures}
    \label{tab:ff4_rates}
    \begin{tabular}{l c c c c}
                        & (1)        & (2)          & (3) \\
                        &  FF4$_t$   & $Monetary_t$ & $News_t$ & Observations \\ \hline \hline
                        &            &              &           &     \\
     Nominal 3 month    & 0.300      &  0.265     & 0.797***  & 106 \\
                        & (0.218)    &  (0.234)   & (0.248)          &  \\
     Nominal 1 year     & 0.486***   &  0.494***  & 0.376     & 106 \\
                        & (0.128)    &  (0.123)   & (0.316)   &  \\    
     Nominal 2 years    & 0.571***   &  0.569***  & 0.594     & 106 \\
                        & (0.209)    &  (0.210)   & (0.509)   &  \\
     Nominal 5 years    & 0.408*     &  0.240     & 0.432     & 106 \\
                        & (0.244)    &  (0.406)  & (0.711)   &  \\
     Nominal 10 years   & 0.149      &  0.143     & 0.235     & 106 \\
                        & (0.178)    &  (0.184)   & (0.593)   &  \\
     Nominal 20 years   & -0.00137   &  0.000402  & -0.0261   & 106 \\
                        & (0.135)    &  (0.138)   & (0.547)   &  \\
     TIPS Real 2 years  & -0.0473    &  -0.0521    & 0.0192  & 74 \\
                        & (0.131)    &  (0.139)    & (0.530)      &  \\
     TIPS Real 5 years  & 0.126      &  0.151      & -0.219  & 82 \\
                        & (0.191)    &  (0.175)    & (0.824) &  \\
     TIPS Real 10 years & 0.0124     &  0.0295     & -0.226  & 82 \\
                        & (0.152)    &  (0.140)    & (0.642) &  \\
     TIPS Real 20 years & 0.0401     &  0.0355     & 0.0954  & 70 \\
                        & (0.162)    &  (0.172)    & (0.587) &  \\
                        &            &             &           &     \\ \hline \hline
    \multicolumn{5}{c}{ Robust Standard Errors in Parentheses} \\
    \multicolumn{5}{c}{ *** p$<$0.01, ** p$<$0.05, * p$<$0.1} \\
    \end{tabular}
\end{table}
Table \ref{tab:ff4_rates} presents the results of this regressions. Column (1) presents the empirical estimates of coefficient $\beta$ in Equation \ref{eq:shock} for different nominal and real rates. Results seem to show a hump-shaped impact of FF4$_t$ on nominal rates, with no significant impact on the three month nominal rate, a positive impact on nominal rates at the one to 10 year maturities (peaking at 2 years), and no impact on long-term rates. Surprisingly, there does not seem to be a significant impact of FF4$_t$ on the daily change of real interest rates captured by TIPS.\footnote{This result could be explained by different reasons. For example, given the policy change on FOMC statements in the year 2000 and the data availability of text, we have a short sample. This problem worsens for TIPS treasuries as they were introduced later in time.}

Next, we turn to estimating the impact of the different isolated components of FF4$_t$ estimated in Section \ref{sec:methodology}. Columns (2) and (3) present empirical estimates of coefficients $\gamma$ and $\mu$ in Equation \ref{eq:shock_decomposition}, respectively, for different nominal and real interest rates. Interestingly, while the "\textit{Monetary}$_t$" component of FF4$_t$ does not seem to have a significant effect on the 3-month maturity treasury, the "\textit{News}$_t$" component has a significant impact on this short run interest rate. The "\textit{News}$_t$" component does not seem to have significant impact on mid and long-term treasury yields. Consequently, the estimated coefficients of the "\textit{Monetary}$_t$" component is similar to the coefficients presented in Column (1) for FF4$_t$.

Second, we estimate the impact of the "Policy News Shocks" or PNS$_t$ constructed by \cite{nakamura2018high} and its decomposition on nominal and real interest rates. Note, this policy indicator is comprised of changes in multiple nominal interest rates at different maturities spanning the first year of the term structure.\footnote{More specifically, the composite policy indicator is the first principal component of the unanticipated change over the 30-minute windows in the following five interest rates: the Fed funds rate immediately following the FOMC meeting, the expected Fed funds rate immediately following the next FOMC meeting, and expected three-month eurodollar interest rates at horizons of two, three, and four quarters.} 
\begin{table}[ht]
    \centering
    \caption{Response of Interest Rates to Policy News Shocks}
    \label{tab:PNS_rates}
    \begin{tabular}{l c c c c}
                        & (1)        & (2)          & (3) \\
                        &  PNS$_t$   & $Monetary_t$ & $News_t$ & Observations \\ \hline \hline
                        &            &              &           &     \\
     Nominal 3 month    & 0.670***   &  0.581***    & 1.208***         & 106 \\
                        & (0.141)    &  (0.150)     & (0.273)          &  \\
     Nominal 1 year     & 0.795***   &  0.834***    & 0.558**          & 106 \\
                        & (0.111)    &  (0.121)     & (0.263)          &  \\     
     Nominal 2 years    & 1.052***   &  1.157***    & 0.420            & 106 \\
                        & (0.203)    &  (0.207)     & (0.394)          &  \\
     Nominal 5 years    & 0.929***   &  1.135***    & -0.310           & 106 \\
                        & (0.226)    &  (0.249)     & (0.581)          &  \\
     Nominal 10 years   & 0.456**    &  0.641***    & -0.655           & 106 \\
                        & (0.184)    &  (0.202)     & (0.516)          &  \\
     Nominal 20 years   & 0.225      &  0.333*      & -0.421           & 106 \\
                        & (0.169)    &  (0.185)     & (0.431)          &  \\
     TIPS Real 2 years  &  1.057***   &  1.129***    & 0.138         & 74 \\
                        & (0.239)    &  (0.241)     & (0.560)          &  \\
     TIPS Real 5 years  & 0.787***   &  0.944***    & -1.729**         & 82 \\
                        & (0.249)    &  (0.268)     & (0.720)          &  \\
     TIPS Real 10 years & 0.543**    &  0.660***    & -1.332*          & 82 \\
                        & (0.210)    &  (0.224)     & (0.673)          &  \\
     TIPS Real 20 years & 0.317      &  0.420*      & -0.850           & 70 \\
                        & (0.210)    &  (0.231)     & (0.643)           &  \\
                        &            &              &           &     \\ \hline \hline
    \multicolumn{5}{c}{ Robust Standard Errors in Parentheses} \\
    \multicolumn{5}{c}{ *** p$<$0.01, ** p$<$0.05, * p$<$0.1} \\
    \end{tabular}
\end{table}
A priori, this could imply greater variability of PNS$_t$, compared to FF4$_t$. Additionally, in \cite{nakamura2018high} the authors stress that this composite policy indicator captures the effects of “forward
guidance” (FOMC announcements that convey information about future changes in the Fed Funds rate). Column (1) of Table \ref{tab:PNS_rates} presents the empirical estimates of coefficient $\beta$ in Equation \ref{eq:shock}. While results are in line with those for FF4$_t$, the estimates are highly more significant. Again, we find that the empirical estimates for $\beta$ exhibits a hump-shaped impact on nominal interest rate across the term structure. Unlike the results presented in \ref{tab:ff4_rates}, the policy news shock has a significant impact on TIPS yields up to the 10 year maturity rate. 

Finally, we turn to estimating the role of the "Monetary$_t$" and "News$_t$" components of Policy New Shocks. Once more, the "News$_t$" component has a large and significant impact on short term nominal interest rates. One way to quantify the relevance of the news component is comparing the estimated coefficients for the "Monetary$_t$" component and those for PNS$_t$. Not controlling for the news component biases up the impact on nominal interest rate between 10 and 5 basis points for short term nominal interest rate, and biases the impact down between 15 and 30 basis points for mid and long term nominal interest rates. In addition, the impact of "Monetary$_t$" on TIPS real rates is greater than the impact of PNS$_t$. Interestingly, the news component which has a large and significant impact on short run nominal interest rate, also seems to have a large but negative impact on long run real rates. 

In summary, the paragraphs above studied the impact of high-frequency financial surprises and its decomposition on the daily change of nominal and real rates around FOMC meetings. We find that the news component of these high-frequency surprises has a significant impact on short term nominal interest rates across different specified shocks. We take these results as evidence that differences in the expected policy rate by the FOMC Statements and the New York Times has a systematic and persistent impact on relevant financial rates. 

\section{Conclusion} \label{sec:conclusion}

We argue that FOMC announcements contain both information and monetary policy content. We develop a transparent framework to pick out the informational content of high-frequency financial surprises around FOMC announcements by using state-of-the art text analysis on both policy statements and newspapers. 

We find that differences in expectations about policy rates between the FOMC and private agents are predictive of high-frequency movements around policy announcements. Furthermore, we show that the isolated informational content has a significant impact on the yields of short-term government bonds. Consequently, ignoring this informational content may bias our understanding of the impact of monetary policy shocks.

We interpret these findings as an existence result. In order to draw reasonable conclusions from a small dataset, we have prioritized simplicity and transparency over more flexible functional forms. Hence, we consider our estimates to be lower bounds for the informational content in FOMC announcements. 


\newpage
\bibliography{main}

\newpage
\appendix

\section{SVAR Evidence} \label{subsec:svar_evidence}

In this Appendix we provide evidence of the informational content in high frequency financial surprises by introducing their isolated components into a SVAR model. In particular, we will focus on estimating the impact of monetary component of the isolated shocks on aggregate variables. 

The most general specification of the SVAR model has $n$ endogenous variables, $p$ lags, and $m$ exogenous variables which can be written in compact form as
\begin{align}
    y_t = A_1 y_{t-1} + \ldots +  A_p y_{t-p} + C x_t + \epsilon_t
\end{align}
where $y_t = \left(y_{1,t}, \ldots, y_{n,t}\right)$ is an $n \times 1$ vector of endogenous variables; $A_1, \ldots, A_p$ are $p$ $n\times n$ matrices; $C$ is a $n\times m$ matrix. The time frequency of the estimated model is monthly. Consequently, we choose the vector $y_t$ to be comprised of three variables: the industrial production growth rate, the consumer price index inflation and the excess bond premium or EBP (see \cite{gilchrist2012credit}). This specification is in line with other empirical models estimated in the literature (see \cite{gertler2015monetary} and \cite{jarocinski2020deconstructing}. In terms of the exogenous vector of variables $m_t$, it will be comprised of the "Monetary$_t$" component of the shocks deconstructed in Section \ref{sec:methodology}. We consider both the isolated component of FF4$_t$ and the PNS$_t$ shocks. We focus on the impact of the "Monetary$_t$" component as the "News$_t$" component does not have a straightforward interpretation for all cases, as stressed in previous sections. Thus, We center our attention on estimating the impact of the component of high-frequency surprises which can not be attributed to differences in expectations between policymakers and the private sector. The model is estimated using Bayesian techniques using a Minnesota prior over its parameters.

\begin{figure}[ht] 
    \centering
    \includegraphics[scale=0.7]{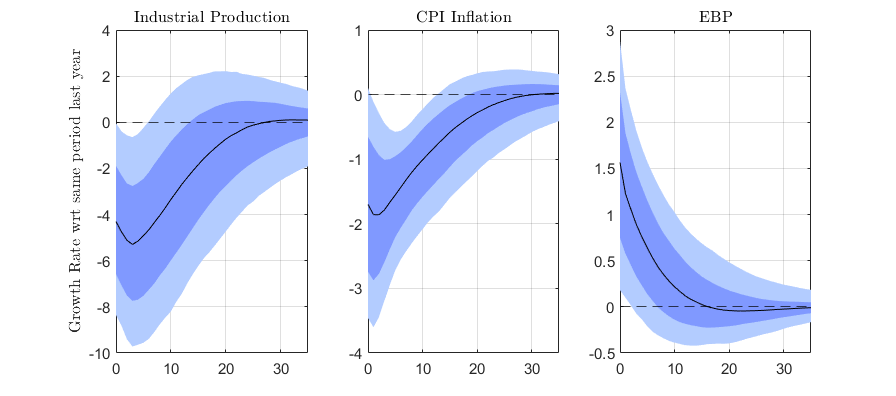}
    \caption{Impulse Response Function - Monetary$_t$ Component \\ Fed Funds Rate Shock}
    \label{fig:IRFs_FF4}
\end{figure}

\begin{figure}[ht] 
    \centering
    \caption{Impulse Response Function - Monetary$_t$ Component \\ Policy News Shock}
    \label{fig:IRFs_PNS}
    \includegraphics[scale=0.7]{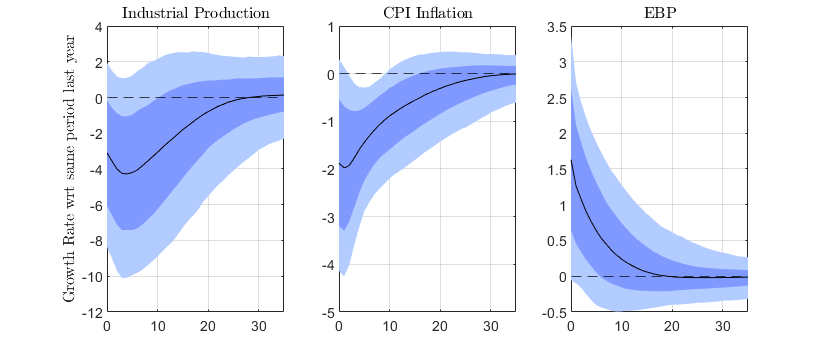}
\end{figure}

Figures \ref{fig:IRFs_FF4} and \ref{fig:IRFs_PNS} present the results of this exercise for the FF4$_t$ and PNS$_t$ shocks respectively. The different shades reflect the 66\% and 90\% confidence intervals.
Across the two shocks, we observe that a one standard deviation increase in the Monetary$_t$ component of the fundamental shocks lead to a drop in the growth rate of industrial production and inflation, and tighter private sector financial conditions. However, the results are statistically different from zero for the FF4$_t$ shock while the results for the PNS$_t$ shock are less clear. This could be due to the PNS$_t$ having a stronger news component, as seen for its impact on daily rates in the previous subsection. Consequently, if the methodology presented in Section \ref{sec:methodology} does not capture the entirety of the informational content, it could be biasing these IRF results.

\end{document}